\documentclass[11pt]{article}
\usepackage{include-file}
\usepackage{authblk}
\usepackage{verbatim}

\def\GD {\textsc{Spectral Graph Dominance~}}

\def\CN {\textsc{Condition-Number~}}
\def \CD {\textsc{Cuts-and-Distances~}}
\def \SRGI {\textsc{Spectrally Robust Graph Isomorphism~}}
\def \qf {{\cal R}}
\def\cR {\mathbb{R}}
\def\cn {\mathbf c}
\def\dl {\mathbf d}

\title{Spectrally Robust Graph Isomorphism}

\author[1]{Alexandra Kolla}
\author[2]{Ioannis Koutis}
\author[3]{Vivek Madan}
\author[4]{Ali Kemal Sinop\footnote{ {\tt alexandra.kolla@colorado.edu}, {\tt ikoutis@njit.edu},  {\tt vmadan2@illinois.edu}, {\tt asinop@gmail.com}. \\Part of this work was supported by NSF grant CCF-1319376.
}}
\affil[1]{Department of Computer Science, University of Colorado at Boulder }
\affil[2]{Department of Computer Science, New Jersey Institute of Technology}
\affil[3]{Department of Computer Science, University of Illinois, Urbana-Champaign}
\affil[4]{TOBB University of Economics and Technology, Ankara}

\date{}

\begin{document}
\maketitle

\begin{abstract}
	\noindent We initiate the study of spectral generalizations of the graph isomorphism problem. 

\smallskip
  \noindent \textbf{(a)} The \emph{Spectral Graph Dominance (SGD)} problem: \\
  On input of two graphs $G$ and $H$ does there exist a permutation
  $\pi$ such that $G\preceq \pi(H)$?
	
\smallskip  	
  \noindent \textbf{(b)} The \emph{Spectrally Robust Graph Isomorphism ($\kappa$-SRGI)} problem: \\
  On input of two graphs $G$ and $H$, find the smallest number $\kappa$ over all permutations $\pi$ such that $  \pi(H) \preceq G\preceq \kappa c \pi(H)$ for some $c$.  SRGI is a natural formulation of the network alignment problem that has various applications, most notably in computational biology.

\smallskip  
  
\noindent  $G\preceq c H$ means that for all vectors $x$ we have $x^T L_G x \leq c x^T L_H x$, where $L_G$ is the Laplacian $G$.
  
\smallskip  

We prove NP-hardness for SGD. 
We also present a $\kappa^3$-approximation algorithm for SRGI for the case when both $G$ and $H$ are bounded-degree trees. The algorithm runs in polynomial time when $\kappa$ is a constant.

  %
  %

\end{abstract}

\section{Introduction}\label{sec:intro}
Network alignment, a problem loosely defined as the comparison of graphs under permutations, has a very long history of applications in disparate fields~\cite{Emmert-Streib:2016}. Notably, alignment of protein and other biological networks are among the most recent and popular applications~\cite{Patrok12,FeiziQMMKJ16}. There are several heuristic algorithms for the problem; naturally some of them are based on generalizations of the graph isomorphism problem, mostly including variants of the robust graph isomorphism problem which asks for a permutation that minimizes the number of `mismatched' edges~\cite{odonnell}.

Robust graph isomorphism may not be always an appropriate problem for applications where one wants to certify the `functional' equivalence of two graphs. Consider for example the case when $G$ and $H$ are two random constant-degree expanders. While they can be arguably functionally equivalent (e.g. as information dispersers), all permutations will incur a large number of edge mismatches, deeming the two graphs very unsimilar.
Functional equivalence is of course an application-dependent notion. In the case of protein networks, it is understood that proteins act as electron carriers~\cite{hanukoglu1996}. Thus it is reasonable to model them as electrical resistive networks that are algebraically captured by graph Laplacian matrices~\cite{doylesnell00}. Going back to the graph isomorphism problem, we note the simple fact that the Laplacian matrices of two isomorphic graphs share the same eigenvalues, with the corresponding eigenspaces being identical up to the isomorphism. We can them aim for a \textbf{spectrally robust} version of graph isomorphism (SRGI) which allows for similar eigenvalues and approximately aligned eigenspaces, up to a permutation. 

In lieu of using directly the eigenvalues and eigenspaces to define SRGI, we will rely on the much cleaner notion of spectral graph similarity, which underlies spectral sparsification of graphs, a notion that has been proven extremely fruitful in algorithm design~\cite{Batson2013, Koutis:2012}. 
More concretely, let us introduce the precise notion of similarity
we will be using. 

\begin{definition} [dominance] \label{def:dominance}
	We say that graph $G$ dominates graph $H$ ($G\preceq H$), when 
	for all vectors $x$, we have $x^T L_G x \leq x^T L_H x$, where $L_G$ is the standard Laplacian matrix for $G$.  
\end{definition}	

\begin{definition} [$\kappa$-similarity] \label{def:similarity}
	We say that graphs $G$ and $H$ are $\kappa$-similar, when
	when there exist numbers $\beta$ and $\gamma$, such that $\kappa=\gamma/\beta$ and 
	$
	        \beta H \preceq G \leq \gamma H.\footnote{Graphs are weighted and $cG$ is graph $G$ with its edge weights multiplied by $c$.}
	$
\end{definition}

We are now ready to introduce our main problem. 

\vspace{2ex}
{\SRGI} ($\kappa$-SRGI): Given two graphs $G,H$, does there exist a permutation $\pi$ on $V(G)$ such that $G$ and $\pi(H)$ are $\kappa$-similar?
\vspace{2ex}

It can been shown that this definition does imply approximately equal eigenvalues and aligned eigenspaces~\cite{Koutis-thesis}, thus testing for $\kappa$-similarity under permutations is indeed a spectrally robust version of graph isomorphism. Going back to our example with the two random expanders, it is well-understood that $G$ and $\pi(H)$ will be $\kappa$-similar for a constant $\kappa$ and for all permutations $\pi$, 
which is what we intuitively expect.

We view spectrally robust graph isomorphism as an interesting theoretical problem due to its close relationship with other fundamental algorithmic questions. In particular, it can be easily seen that $\kappa$-SRGI is equivalent to the graph isomorphism problem when $\kappa=1$. As we will discuss in more detail, SRGI can also be viewed as a natural generalization of the minimum distortion problem~\cite{KRS}. 
Up to our knowledge, the spectral-similarity approach to network alignment has been mentioned earlier only in~\cite{Tsourakakis14}. 
In view of the vast number of works on GI (\cite{GI1,GI2, GI3,
	GI4,GI5,luks,babai} to mention a few) as well as the works on the
robust graph isomorphism problem \cite{odonnell} and the minimum
distortion problem \cite{KRS}, we find it surprising that SRGI
has not received a wider attention.

The goal of this work is to prove some initial results on SRGI and stimulate further research. Towards that end, we provide the first
algorithm for this problem, for the case when both graphs are trees. 

\begin{theorem} \label{thm:main} 
	Given two $\kappa$-similar trees $G$ and $H$ of maximum degree
	$d$, there exists an algorithm running in time $O(n^{O(k^2d)})$ which finds a mapping certifying that the they are at most $\kappa^4$-similar. 
\end{theorem}

The algorithm for trees is already highly involved, which gives grounds for speculating that the problem is NP-hard. We give evidence that this may be indeed true by turning our attention to the one-sided version of the problem. 

\vspace{2ex}
{\GD} (SGD): Given two graphs $G,H$, does there exist a permutation $\pi$ such that $G$ dominates $\pi(H)$? 
\vspace{2ex}

Given two graphs $G$ and $H$ that have the same eigenvalues, it is not hard to prove
that if $G$ and $H$ are not isomorphic, then $G$ cannot dominate $H$ (and vice-versa).
Combining this with the fact that isomorphic graphs have the same eigenvalues, 
we infer that SGD is at least graph isomorphism-hard. The second contribution of this work is the following theorem.

\begin{theorem}\label{thm:np-hardness}
	The \GD problem is NP-hard.
\end{theorem}

Theorem~\ref{thm:np-hardness} is proved in Section~\ref{sec:gd}. We can actually prove a slightly stronger theorem that restricts one of the input graphs to be a tree.

\subsection{Related Work}
The Robust Graph Isomorphism problem (RGI) asks for a permutation
that minimizes the number of mismatched edges. 
O'Donnell~{\it et al.}~\cite{odonnell} gave a constant
factor hardness for RGI . The Minimum Distortion problem (MD) views graphs as distance metrics, using
the shortest path metric. The goal is to find a mapping between
the two metrics so as to minimize the maximum
distortion. The connection
between SRGI and MD stems from the observation that if two tree graphs
$G$ and $H$ are $\kappa$-similar up to a permutation
$\pi$, then the distortion between the
induced graph distances of $G$ and $\pi(H)$ is at most $\kappa$. 
For the MD problem, Kenyon {\it et al.}~\cite{KRS} gave an
algorithm which finds a solution with distortion at most $\alpha$
(provided that it exists) in time
$\mathrm{poly} (n) \exp(d^{O(\alpha^3)})$, for a tree of degree
at most $g$ and an arbitrary weighted graph. They also prove that this problem is NP-hard to
approximate within a constant factor.

The term `spectral alignment' has been used before in~\cite{FeiziQMMKJ16} in the context of spectral relaxation of the graph matching function. The algorithm in~\cite{Patrok12} is more spectral `in spirit' because it
uses directly the spectral of the normalized Laplacians of several subgraphs to construct complicated `graph signatures' that are then compared for similarity. There is no underlying objective function that drives 
the computation of these signatures, but we imagine that the proposed algorithm or some variant 
of it, may be a reasonably good practical candidate for SRGI.
The work by Tsourakakis~\cite{Tsourakakis14} proposes an algorithm that searches
for the optimal permutation via a sequence of transpositions; however
the running time of the algorithm does not have any non-trivial sub-exponential
upper bound.

\section{Graph Dominance}\label{sec:gd}
\noindent \textbf{Preliminaries}. Given a weighted graph $G=(V,E,w)$ we denote by $E_G$ its edges. The Laplacian $L_G$ of $G$ is the matrix defined by $L(i,j)=-w_{ij}$ and $L(i,i) = \sum_{i\neq j} w_{ij}$. 
The quadratic form $\qf(G,x)$  of $G$ is the function defined as:
\begin{equation} \label{eq:quadratic}
\qf(G,x) = x^T L_G x = \sum_{i,j} w_{ij}(x_i-x_j)^2.
\end{equation}
Let $G^{\infty}$ be the infinite graph with vertex set equal to all points on the plane with integer coordinates. There is an edge between two points of $G^{\infty}$ if they have Euclidean distance one. A \textit{cubic subgrid} is a finite subgraph of $G^{\infty}$ such that all of its nodes have degree at most $3$. 


The main ingredient of the proof is the following theorem. 

\begin{theorem}\label{thm:hamiltonian_spectral}
Let $G$ be a cubic subgrid and $C$ be the cycle graph, both on $n$ vertices. There exists
a permutation $\pi$ such that $\pi(C) \preceq G$ if and only if $G$ contains a Hamiltonian cycle. 
\end{theorem}
\begin{proof}  If $G$ contains a Hamiltonian cycle $\pi(C)$, then equation~\ref{eq:quadratic} directly implies that $\pi(C)\preceq G$.  To prove the converse  assume that $G$ does not contain a Hamiltonian cycle and let $H$ be a permutation of $C$ such that $|E_G\cap E_H|$ is maximized. We prove a number of claims and lemmas.

\begin{claim}\label{claim:deleting_shared_edges}
Let $G', H'$ be the graphs obtained by deleting the common edges between $G$ and $H$ respectively. Then, $\qf(G,x)< \qf(H,x)$ if and only if $\qf(G',x) < \qf(H',x)$.
\end{claim}

\begin{proof}
Let $F$ be the graph induced by the edges shared by $G$ and $H$. By equation~\ref{eq:quadratic} we have
$\qf(G,x) = \qf(G',x)+ \qf(F,x)$ and $ \qf(H,x) = \qf(H',x)+ \qf(F,x)$. The claim follows.
\end{proof}

\begin{claim}\label{claim:deleting_degree_1_edge}
Let $v$ be a vertex with $deg_{G'}(v)=1$, $deg_{H'}(v)=0$ and let $G''$ be the graph obtained from $G'$ after deleting the edge incident to $v$, and set $H''=H'$. Then, there exists a vector $x$ s.t. $\qf(H',x)>\qf(G',x)$ iff there exists a vector $y$ s.t. $\qf(H'',y)>\qf(G'',y)$.
\end{claim}
\begin{proof}
Let $x$ be a vector such that $\qf(H',x)>\qf(G',x)$. Since $G''$ is a subgraph of $G'$, we have $\qf(G'',x) \leq \qf(G',x) < \qf(H',x) = \qf(H'',x) $, and we can take $y =x$. For the converse, assume that there is a vector $y$ such that $\qf(H'',y)>\qf(G'',y)$. Let $(v,w)$ be the edge incident to $v$ in $G'$. We define a vector $x$ as follows: $x_{u} = y_u$ for all $u \neq v$ and $x_v = y_w$. Since, $deg_{H''}(v) = 0$, we have 
$\qf(H'',y) = \qf(H'',x)$. On the other hand,  $G'$ and $G''$ agree on all the edges except $(v,w)$. Hence, 
$\qf(G',x) = \qf(G'',x) + (x_v - x_w)^2 = \qf(G'',x) $. The two vectors $x$ and $y$ agree on all the entries except at $v$, and the degree of $v$ in $G''$ is zero. Hence, $\qf(G'',x) = \qf(G'',y)$. Combining all the inequalities, we get: \\
$ \qf(G',x)= \qf(G'',x) = \qf(G'',x) < \qf(H'',y) = \qf(H'',x) = \qf(H',x)$.
\end{proof}

\begin{claim}\label{claim:degree_G_H}
Let $G'$ and $H'$ be the graphs obtained by deleting the shared edges between $G$ and $H$ as in
Claim~\ref{claim:deleting_shared_edges}.
Let $\tilde{G}$ and $\tilde{H}$ be the graphs obtained starting from $G'$ and $H'$ and repeatedly
applying the edge deletion operation of Claim~\ref{claim:deleting_degree_1_edge}. Then, for any vertex $u$, $deg_{\tilde{G}}(u) \leq deg_{\tilde{H}}(u) + 1$.  
\end{claim}
\begin{proof}
Since $G$ is a cubic subgrid graph and $H$ is a cycle, $ deg_G(u) \leq 3$, $deg_H(u) = 2$, for all vertices $u$. Deleting edges shared between $G$ and $H$ decreases the degree of any given vertex by the same amount in $G$ and $H$. Moreover, at any given step, we only delete edges from $G'$. Hence, $deg_{\tilde{G}}(u) \leq deg_{\tilde{H}}(u) + 1$. 
\end{proof}

\begin{claim}\label{claim:degree_1_G_H} 
Let $G'$ and $H'$ be the graphs obtained by deleting the shared edges between $G$ and $H$ as in
Claim~\ref{claim:deleting_shared_edges}. If there exists a vertex $v$ such that $deg_{G'}(v) = 1, deg_{H'}(v) \geq 1$. Then, there exists a vector $x$ such that $\qf(H',x) > \qf(G',x)$.
\end{claim}
\begin{proof} Let the edge incident to $v$ in $G'$ be $(v,w)$ and an edge incident to $v$ in $H'$ be $(v,u)$. Since, $H'$ and $G'$ do not share any edge, we have $u \neq w$. Let $x \in \cR^n$ be a vector defined as follows: $x_v = 0, x_w = \frac{1}{2}$ and  $x_t = 1$ otherwise. We have $
\qf(H',x) > (x_v-x_u)^2 = 1,
$
and
\begin{equation*}
\qf(G',x) = (x_v - x_w)^2 + \sum_{(w,a) \in E_G', a \neq v} (x_w-x_a)^2
\end{equation*}
Vertex $w$ has at most two neighbors other than $v$  in $G$, since $deg_{G'}(v) \leq 3$ and for any such neighbor $a$, we have $(x_a-x_w)^2 = (\frac{1}{2}-1)^2 = 1/4$. Hence $\qf(G',x) \leq 3/4 <1 \leq \qf(H',x)$.
\end{proof}

Let $G'$ and $H'$ be the graphs obtained by deleting the shared edges between $G$ and $H$ as in
Claim~\ref{claim:deleting_shared_edges}. Claims~\ref{claim:degree_G_H} and~\ref{claim:degree_1_G_H}
allow us to assume without loss of generality that there is no degree one vertex in $G'$ and for all vertices $u$, $deg_{G'}(u) \leq deg_{H'}(u) + 1$. For 
convenience, we will refer to the edges of $G'$ as \textbf{black edges} and edges of $H'$ as \textbf{blue edges}.

\begin{lemma}\label{lem:w2_degree3}
If there exist five vertices $u,v,w_1,w_2,w_3$ such that 
\begin{itemize}
\item $(u,w_1),(w_1,w_2),(w_2,w_3)$ are black edges and $(v,w_1), (v,w_2)$ are \textbf{not} black edges.
\item $(u,v)$ is a blue edge and $(u,w_2)$ is \textbf{not} a blue edge.
\end{itemize}
Then, there exists a vector $x$ such that $\qf(H',x)> \qf(G',x).$
\end{lemma}
\begin{proof} Let $x$ be the vector with $x_u=0$, $x_v$=2, $x_{w_1}=\frac{1}{3}$ and $x_{w_2}=\frac{2}{3}$, and $x_t=1$ otherwise. 
We have
\begin{eqnarray*}
\qf(H',x)  & \geq & (x_u - x_v)^2 + \sum_{\substack{(u,a) \in E_{H'}\\ a \neq v}} (x_u - x_a)^2 + \sum_{\substack{(v,b)\in E_{H'} \\ b \neq u}} (x_v - x_b)^2\\
& = & 4 + (deg_{H'}(u) - 1)\cdot(0-1)^2  +   (deg_{H'}(v) - 1) \cdot (2-1)^2 \\ 
& = & deg_{H'}(u) + deg_{H'}(v) + 2 \geq deg_{G'}(u) + deg_{G'}(v) \mbox{  \qquad      (Claim~\ref{claim:degree_G_H})}
\end{eqnarray*}
and
\begin{eqnarray*}
\qf(G',x)  & = &  (x_u - x_{w_1})^2 + (x_{w_1} - x_{w_2})^2 + (x_{w_2} - x_{w_3})^2+ \sum_{\substack{(u,a) \in E_{G'}\\ a \neq w_1}} (x_u - x_a)^2 \\
          & &   + \sum_{\substack{(w_1,c) \in E_{G'}  \\ c \neq w_2,u}} (x_{w_1} - x_c)^2 + \sum_{(v,b) \in E_{G'}} (x_v - x_b)^2 + \sum_{\substack{(w_2,d) \in E_{G'}\\ d \neq w_1,w_3}}(x_{w_2} - x_d)^2.
\end{eqnarray*}

We observe that \textbf{(i)}~The first three terms are equal to $\frac{1}{9}$. \textbf{(ii)}~There is at most one edge $(w_1,c)$ for $c\neq w_2,u$. Also, since $w_1$ is not incident to $v$, we have $x_c=1$. Thus the fifth term is at most $\frac{4}{9}$. \textbf{(iii)}  There is at most one edge $(w_2,d)$ for $d\neq w_1,w_3$. In addition, $G'$ is a subgrid, so there is no cycle of length 3 and $w_2$ is not incident to $u$. Also, $w_2$ is not incident to $v$, by assumption. So, it must be that $x_d=1$ and the last term is at most equal to $\frac{1}{9}$. \textbf{(iv)}~Since $G'$ and $H'$ do not share an edge, $u$ is not connected to $v$. By assumption $u$ is also not incident to $w_2$. So, it must be that $x_a=1$ and the fourth term is equal to $deg_{G'}(u)-1$. \textbf{(v)}~Vertex $v$ is not connected to $u, w_1, w_2$. Thus it must be $x_b=1$ and the sixth term is equal to $deg_{G'}(v)$. 
Collecting the terms gives $\qf(G
,x)\leq deg_{G'}(u)+ deg_{G'}(v) -\frac{1}{9}$ and the Lemma follows. 
\end{proof}
\begin{lemma}\label{lem:w2_degree2}
If there exist four different vertices $u,v,w_1,w_2$ such that 
\begin{itemize}
\item $w_1$ has only two black adjacent edges $(u,w_1)$ and $(w_1,w_2)$ 
\item $(u,v)$ is a blue edge.
\end{itemize}
Then, there exists a vector $x$ such that $\qf(H',x)> \qf(G',x).$
\end{lemma}
\begin{proof}
Let $x$ be a vector with $ x_u = 0, x_v = 2, x_{w_1} = \frac{1}{2}$ and $x_t=1$ otherwise. 
We have
\begin{eqnarray*}
\qf(H',x)  & \geq & (x_u - x_v)^2 + \sum_{\substack{(u,a) \in E_{H'}\\ a \neq v}} (x_u - x_a)^2 + 
\sum_{\substack{(v,b)\in E_{H'}\\ b \neq u}} (x_v - x_b)^2\\
& = & 4 + (deg_{H'}(u) - 1) \cdot (0-1)^2 + (deg_{H'}(v) - 1) \cdot (2-1)^2   \mbox{\qquad (no shared edges)}\\
& = & deg_{H'}(u) + deg_{H'}(v) + 2 \geq deg_{G'}(u) + deg_{G'}(v) \mbox{\qquad (Claim~\ref{claim:degree_G_H})}
\end{eqnarray*}
and
\begin{eqnarray*}
\qf(G',x) & = &  (x_u - x_{w_1})^2 + (x_{w_1} - x_{w_2})^2 + \sum_{\substack{(u,a) \in E_{G'}\\ a \neq w_1}} (x_u - x_a)^2 + \sum_{(v,b) \in E_{G'}} (x_v - x_b)^2 
\end{eqnarray*}

Since $G'$ and $H'$ do not share an edge, $(x_u - x_a)^2$ and $(x_v-x_b)^2$ terms are $(0-1)^2$ and $(2-1)^2$ respectively. We have
\begin{eqnarray*}
\qf(G',x) &=& \frac{1}{4} + \frac{1}{4} + (deg_{G'}(u) - 1) \cdot 1 + deg_{G'}(v) \cdot 1 = deg_{G'}(u) + deg_{G'}(v) -\frac{1}{2}.
\end{eqnarray*}
The Lemma follows. 
\end{proof}

\begin{lemma}\label{lem:vertex_degree_3}
If there exists a degree three vertex in $G'$, then there exists a vector $x$ such that $\qf(H',x)> \qf(G',x).$ 
\end{lemma}
\begin{proof}
Since $deg_{G'}(u) = 3, deg_{H'}(u) \geq 2$ by claim~\ref{claim:degree_G_H}. Consider the underlying grid of which $G$ is a subgraph. Pick the edge $(u,v) \in E_{H^{'}}$ which is either not axis-parallel or is axis-parallel and $v$ is at distance at least $2$ in the grid. Since $u$ has degree $3$ in $G'$, there exists a neighbor $w_1$ of $u$ in $G'$ such that any path from $w_1$ to $v$ in $G'$ has length at least $3$. 

If $w_1$ has degree $2$ in $G'$, then we set $w_2$ to be the neighbor of $w_1$ other than $u$. It is then straightforward to check that $u,v,w_1,w_2$ satisfy the condition of Lemma~\ref{lem:w2_degree2}. Hence, there exists $x$ such that $\qf(H',x)> \qf(G',x).$

If $w_1$ is not of degree $2$ in $G'$, it must have degree $3$ since there are no degree $1$ vertices in $G'$. Let $w_2$ be the neighbor of $w_1$ other than $u$ such that $(u,w_2) \not \in E_H'$. Such a neighbor exists since $u$ has at most one neighbor in $H'$ other than $v$ and $v$ is not incident to $w_1$ due to the fact that any path of length from $w_1$ to $v$ has length at least $3$. Let $w_3$ be the neighbor of $w_2$ other than $w_1$. Such a neighbor must exist since there is no vertex of degree $1$ in $G'$. Now, we prove that these vertices satisfy the condition of Lemma~\ref{lem:w2_degree3}. By construction $(u,w_1),(w_1,w_2),(w_2,w_3)$ are black eges. Any path from $w_1$ to $v$ in $G'$ has length at least $3$ which implies that $(w_1,v),(w_2,v)$ are not black edges. Also, by construction, $(u,v)$ is a blue edge and $(u,w_2)$ is not a blue edge. Hence, by lemma~\ref{lem:w2_degree3} there exists $x$ such that $\qf(H',x)> \qf(G',x).$ 
\end{proof}

\begin{lemma}\label{lem:cut_vector}
If there exists a set $S$ of vertices such that no edges leave $S$ in $G'$, but at least one edge leaves $S$ in $H'$ then then there exists a vector $x \in \cR^n$ such that $\qf(H',x)> \qf(G',x).$
\end{lemma}
\begin{proof}
Let $x$ be defined as follows: $x_u = 1$ for $u \in S$ and $x_u = 0$ for $u \not \in S$. The $\qf(G',x)$ is equal to the number of edges leaving $S$ in $G'$, and similarly for $H'$. The lemma follows. 
\end{proof}

\begin{lemma}\label{lem:cycle_length_5}
If there exists a cycle of length more than 4 in $G'$ such that all vertices of cycle have degree $2$ in $G'$, then there exists $x \in \cR^n$ such that $\qf(H',x)> \qf(G',x).$
\end{lemma}
\begin{proof}
Let $C$ be the set of vertices in the cycle. For any vertex $v$ in $C$, $deg_{G'}(v) = 2$. By claim~\ref{claim:degree_G_H}, $deg_{H'}(v) \geq 1$. If there is no blue edge connecting two vertices of $C$ in $G'$, then there are at least $|C|$ edges going out of $C$ in $H'$ and no edge going out of $C$ in $G'$. Then, by lemma~\ref{lem:cut_vector}, there exists $x$ such that $\qf(H',x)> \qf(G',x)$. 

In the complementary case, suppose there is an edge $(a,b) \in E_{H'}$ such that $a,b \in C$. Let the two paths from $a$ to $b$ on $C$ be $P_1 = a,w_1,\dots,w_{k_1},b$ and $P_2 = a,v_1,\dots,v_{k_2},b$. Since $G'$ and $H'$ do not share any edge, $\min(k_1,k_2) \geq 1$. And since the cycle has length at least $5$, $\max(k_1,k_2) \geq 2$. Let a vector $x$ be defined as follows: $x_{a} = 0, x_b = 1, x_{w_i} = \frac{i}{k_1+1}, x_{v_i} = \frac{i}{k_2+1}$ and for $u \not \in C$, set $x_u = 0$. We have

\begin{eqnarray*}
\qf(G',x) & = & \sum_{(u,v) \in P_1} (x_u - x_v)^2 + \sum_{(u,v) \in P_2} (x_u-x_v)^2 \\
& = & (k_1+1)\cdot \frac{1}{(k_1+1)^2} + (k_2+1)\cdot \frac{1}{(k_2+1)^2} \leq  \frac{5}{6}.
\end{eqnarray*}
The last inequality holds because $\min(k_1,k_2) \geq 1, \max(k_1,k_2) \geq 2$. \\ On the other hand
$\qf(H',x) \geq (x_a-x_b)^2 =1 $ and the lemma follows. 
\end{proof}


\begin{lemma}\label{lem:cycles_length_4}
If $G'$ contains a set of disjoint cycles of length $4$ and $H'$ edges only have endpoints on the same cycle, then there is a cycle $\tilde{H}$ such that $|E_{\tilde{H}} \cap E_G| = |E_H \cap E_G|+2$.
\end{lemma}
\begin{proof}
 Consider a cycle $C$ of length $4$ in $G'$ such that there is no blue edge between a vertex in the cycle and a vertex not in the cycle. Since $deg_{G'}(v) \geq 2$ for all vertices in the cycle, $deg_{H'}(v) \geq 1$ by~\cref{claim:degree_G_H}. Since $G'$ and $H'$ do not share any edge and the cycle has length $4$, we must have $deg_{H'}(v) = 1$ for all $v \in C$. And for vertices not in the length $4$ cycles, $deg_{H'}(v) = 0$. Let the edges of $H$ be $F_1 \cup F_2$ where $F_1$ are the edges shared between $G$ and $H$ and $F_2$ are the diagonal edges in the disjoint cycles of length $4$. Let $C$ be one such cycle in $G'$ with vertices $v_1,v_2,v_3,v_4$ in this order and $(v_1,v_3) \in F_2, (v_2,v_4) \in F_2$. Let $H_1 = (V, F_1 \cup F_2 \setminus \{(v_1,v_3), (v_2,v_4)\}\cup \{(v_1,v_2),(v_3,v_4)\}), H_2 = (V,F_1 \cup F_2 \setminus \{(v_1,v_3), (v_2,v_4)\}\cup \{(v_1,v_4),(v_2,v_3)\})$. Then, one of the $H_1$ or $H_2$ is a cycle of length $n$. We let $\tilde{H}$ be that cycle. 
\end{proof}

\noindent {\bf Finishing the proof:}  Recall that we have assumed that $G$ does not contain a Hamiltonian cycle and let $H$ is a permutation of $C$ such that $|E_G\cap E_H|$ is maximized. To show that $G$ does \textbf{not} dominate $H$, we need to construct a vector $x$ such that $\qf(H,x)>\qf(G,x)$. 

Starting with $G$ and $H$, we form two graphs $G'$ and $H'$ as follows: (i) delete from $G$ and $H$ all common edges, (ii) iteratively and greedily delete all vertices such that $deg_{G}(u) = 1, deg_{H}(u) = 0$. Then Claims~\ref{claim:deleting_shared_edges} and \ref{claim:deleting_degree_1_edge} show that it suffices to find a vector $x$ such that $\qf(H',x)>\qf(G',x)$. 

If there is still a vertex with $deg_{G'}(u) = 1$, then $deg_{H'}(u)$ must be at least $1$ and hence, by Claim~\ref{claim:degree_1_G_H}, there exists a vector $x$ such that $\qf(H',x) > \qf(G',x)$. 
Also, if there is a vertex $u$ with degree $3$ in $G'$, then by lemma~\ref{lem:vertex_degree_3} there exists $x$ such that $\qf(H',x)>\qf(G',x)$.

If there are no degree $1$ or degree $3$ vertices $G'$, then $G'$ must be a collection of isolated vertices and cycles. If there is a vertex $v$ such that $deg_{G'}(v) = 0, deg_{H'}(v) \geq 1$, then by setting $S = \{v\}$, lemma~\ref{lem:cut_vector} implies that there exists a vector $x$ such that $\qf(H',x)>\qf(G',x)$.

So, if none of the above cases occurs, then $G'$ is a collection of disjoint cycles and $H'$ edges are only incident to vertices of the cycles. If there is a cycle of length at least $5$, then by lemma~\ref{lem:cycle_length_5} there exists $x$ such that $\qf(H',x)>\qf(G',x)$. Otherwise, if there is at least one blue edge with end points on two different cycles of length $4$, then by setting $S$ to be the vertex set of the cycle of length $4$ Lemma~\ref{lem:cut_vector} implies that there exists $x$ such that $\qf(H',x)>\qf(G',x)$. 

So, either $G'$ and $H'$ are empty or $G'$ consists of a collection of disjoint cycles of length $4$ such that blue edges have end points in the same cycle. In the first case, $G$ trivially contains a Hamiltonian cycle since $H'$ is empty. This is a contradiction to the assumption that $G$ does not contain a Hamiltonian cycle. 
In the second case $G$ Lemma~\ref{lem:cycles_length_4} contradicts our assumption about the maximality
of $|E_G\cap E_H|$. 
\end{proof}

\begin{proof}(Theorem~\ref{thm:np-hardness})
The problem of detecting if a cubic subgrid contains a Hamiltonian cycle is NP-complete~\cite{PapadimitriouV84}.
Hence Theorem~\ref{thm:hamiltonian_spectral}  is a direct reduction, and the theorem follows. 
\end{proof}

\medskip

\section{Spectrally Robust Graph Isomorphism on Trees}\label{subsec:algorithm}
This section outlines a proof for Theorem~\ref{thm:main}. We first introduce necessary notation.

\begin{definition}
	The support $\sigma(G,H)$ of $G$ by $H$ is the smallest number 
	$\gamma$ such that $\gamma H$ dominates $G$. 
\end{definition}

\begin{definition}
	The condition $\kappa(G,H)$ of a pair of graphs $G$ and $H$
	is the smallest number $\kappa$ such that $G$ and $H$ are $\kappa$-similar. We have $\kappa(G,H) = \sigma(G,H) \sigma(H,G)$.	
\end{definition}

We denote by $d_G(u,v)$ the distance between $u$ and $v$ in $G$ using
the shortest path metric. For $S\subseteq V$, we denote by
$\delta_G(S)$ the set of of edges crossing the cut $(S,V-S)$
in $G$.

We now briefly review well-known facts~\cite{GuatteryM00};
a more detailed version of this paragraph along with proofs can be
found in the appendix of the full version of the paper. 
Given two trees $G$ and $H$ there is an obvious way to embed
the edges of $G$ into $H$: each edge $(u,v)$ is routed over
the unique path between vertices $(u,v)$ in $H$. The \emph{dilation}
of the embedding is defined by: 
$
     {\mathbf d} = \max_{(u,v)\in E_G} d_H(u,v).
$
The congestion $c_e$ of an edge $e\in E_H$ is the number
of $G$-edges that are routed over $e$. The \emph{congestion} ${\mathbf c}$
of the embedding is then defined as $\max_{e\in E_H} c_e$. 
An upper bound of $\kappa$ on the condition number
implies the same upper bound on both $c$ and $d$. On the other hand, the product $\cn \dl$
is an upper bound on $\sigma(G,H)$, which is at most a quadratic
over-estimation of $\sigma(G,H)$.

Our algorithm finds a mapping  that controls both the dilation and the congestion of the embeddings from $G$ to $H$ and vice versa, thus obtaining a quadratic approximation to the condition number as a corollary.

\smallskip
\noindent \textbf{Remark:} To simplify the presentation and the proof, we assume uniform upper bounds on the congestion and the dilation of both embeddings ($G$ to $H$ and $H$ to $G$), rather than handling them separately. This formally proves a $\kappa^3$-approximation to $\kappa$-similarity. 
We omit a $\kappa$-approximation algorithm for the final version of the paper. 

Formally, our result can be stated as follows:

\begin{theorem}\label{thm:cuts_distances_algo} Suppose $G$ and $H$ are two trees 
  for which there exists a bijective mapping
  $\pi:V(G) \rightarrow V(H)$ satisfying the following properties:
  \begin{itemize}
  \item For all $(u,v) \in E(G)$, $d_H(\pi(u),\pi(v)) \leq \ell$
  \item For all $(u,v) \in E(H)$, $d_G(\pi^{-1}(u),\pi^{-1}(v)) \leq \ell$
  \item For $S \subset V(G)$ such that $|\delta_G(S)| = 1$, $|\delta_H(\pi(S))| \leq k$.
  \item For $S \subset V(H)$ such that $|\delta_H(S)| = 1$, $|\delta_G(\pi^{-1}(S))| \leq k$.
  \end{itemize}
  Then, there exists an algorithm to find such a mapping in time
  $n^{O(k^2d)}$ where $d$ is the maximum degree of a vertex in $G$ or
  $H$.
\end{theorem}

Our main result, theorem \ref{thm:main}, follows immediately as a corollary from the fact that
$\max\{k,l\} \leq \sigma(H,G) \leq kl$ (which is proved in the full version of the paper) and 
using the fact that $\max \{\sigma(G,H), \sigma(H,G) \} \leq \kappa(G,H) \leq \sigma(G,H) \sigma(H,G)$.

%
\begin{corollary}
  Given two tree graphs $G$ and $H$ with condition number $\kappa$ and maximum  degree $d$, there exists an algorithm running in time  $n^{O(\kappa^2d)}$ which finds a mapping certifying that condition
  number is at most $\kappa^4$.
\end{corollary}
The algorithm uses dynamic programming; it proceeds by recursively finding mappings for different subtrees and merging them. The challenge is to find partial mappings of subtrees which also map their boundaries in such a way that enables different mappings to be appropriately merged. Notice that it is not enough to consider just the boundary vertices of the subtrees and their images. Instead, we need to additionally consider the boundary edges of those vertex sets, which correspond to cuts induced on the graph. 

\medskip

\noindent \textbf{\large Definitions and Lemmas.} To proceed with the proof, we introduce some definitions. We fix $k$ and $\ell$ to be defined as in Theorem~\ref{thm:cuts_distances_algo}. Also, we fix an arbitrary ordering $L$ on the edges of $H$. Without loss of generality it is convenient to root the trees such that we always map the two roots to each other. Let $r_G$ be the root of tree $G$ and $r_H$ be the root of tree $H$.

Suppose that $u$ is a vertex in $G$ and $T^G_u$ is the subtree rooted at $u$ in $G$. If $T_u^G$ is mapped to the set $T$ in $H$, then its boundary includes the vertex $u$, the edge
incident to $u$, the boundary vertices of $T$, and the cuts induced by
edges going out of $T$. Hence, in addition to considering the mapping
of boundary vertices, we need to consider the mapping of sets $T'$
such that $\delta_H(T') = \{e\}$ where $e \in \delta_H(T)$. This
notion is formalized in the following two definitions. 

\begin{definition}\label{def:gamma}
  Let $\Gamma$ be the set of tuples
  $(u,T,v,u_1,\dots,u_x,S_1,\dots,S_x)$ satisfying the following
  properties:
  \begin{itemize}
  \item
    $u, u_1, \dots, u_x \in V(G), v \in V(H), T \subset V(H),
    S_1, \dots, S_x \subset V(G)$;
  \item $r_G \not \in S_1,\dots,S_x, r_H \not \in T$;
  \item $u,u_1,\dots,u_x \neq r_G, v \neq r_H$;
  \item $|\delta_H(T)| = x \leq k$ and
    $\forall j \in [1,x], |\delta_G(S_j)| \leq k $.
  \end{itemize}
\end{definition}

For $\alpha \in \Gamma$, we use the indicator variable $z_{\alpha}$ to
denote
if there is a mapping $\pi$ which realizes $\alpha$ and preserves the
distances and cuts for edges in $T_u^G$ and $T$. A permutation $\pi$ realizing $\alpha$ is formally defined below. Intuitively, this
mapping maps the subtree rooted at $u$ in $G$ to the set $T$ in $H$,
vertex $u$ to vertex $v$. It also maps $u_1,\dots,u_x$ to the vertex
boundary of the set $T$, and maps sets $S_1,\dots,S_x$ to the cuts
induced by the boundary edges of $T$. The formal definition of
$z_\alpha$ is as follows:
\begin{definition}
  \label{def:z_alpha}
  For $\alpha=(u,T,v,u_1,\dots,u_x,S_1,\dots,S_x)\in \Gamma$, let
  $\delta_H(T) = \{e_1,\dots,e_x\}$ be such that for $i<j$, $e_i$ is
  ordered before $e_j$ in ordering $L$. Let $v_j = e_j \cap T$,
  $T_u^G$ be the vertex set in the subtree rooted at $u$ in $G$ and
  for $e \in E(G)$, let $T_e^G$ be the vertex set in the subtree under
  edge $e$. Formally speaking $T_e^G \subset V(G)$ such that
  $\delta_G(T_e^G)= \{e\}$ and $r_G \not \in T_e^G$ ($T_e^H$ is similarly
  defined). We define $z_\alpha= 1$ if there exists a mapping
  $\pi:V(G) \rightarrow V(H)$ such that:
  \begin{enumerate}
  \item \label{item:def-z-alpha-1}
    $\pi(T_u^G) = T, \pi(u) = v, \forall j \in [1,x], \pi(u_j) = v_j,
    \pi(S_j) = T_{e_j}^H$, $\pi(r_G) = r_H$.
  \item \label{item:def-z-alpha-2}
    $\forall (u,v) \in E[G[T_u^G]], d_H(\pi(u),\pi(v)) \leq \ell$
  \item \label{item:def-z-alpha-3}
    $\forall (u,v) \in E[H[T]], d_G(\pi^{-1}(u),\pi^{-1}(v)) \leq
    \ell$
  \item \label{item:def-z-alpha-4}
    $\forall e \in E[G[T_u^G]]$, $|\delta_H(\pi(T_e^G)| \leq k$.
  \item \label{item:def-z-alpha-5}
    $\forall e \in E[H[T]]$, $|\delta_G(\pi^{-1}(T_e^H)| \leq k$.
  \end{enumerate}
  We refer to such a mapping $\pi$ as a certificate of $z_\alpha =
  1$. Moreover, we define $z_{\alpha,\pi} = 1$ if $\pi$ is a
  certificate of $z_{\alpha} = 1$ and $0$ otherwise.
\end{definition}

\begin{claim}\label{prop:z_alpha_pi}
  There exists a $poly(n)$ time algorithm which given
  $\alpha \in \Gamma, \pi:V(G) \rightarrow V(H)$, outputs the value of
  $z_{\alpha,\pi}$.
\end{claim}

Our goal is to design an algorithm which computes $z_{\alpha}$ for
every $\alpha \in \Gamma$. However, for our algorithm to run in
polynomial time, we need $\Gamma$ to not be exponentially large.

\begin{lemma}  $|\Gamma| \leq n^{O(k^2)}$.
\end{lemma}
\begin{proof}
  Let $\alpha=(u,T,v,u_1,\dots,u_x,S_1,\dots,S_x)$. We prove the lemma
  by bounding the number of choices for each parameter.
  \begin{itemize}
  \item The number of choices of $u$ is upper bounded by $n$.
  \item Since $|\delta_H(T)|=x$, the number of choices $\delta_H(T)$
    is upper bounded by ${m \choose x}$ where $m$ is the number of
    edges. By substituting $m = n-1$, we get that the number of
    different $\delta_H(T)$, i.e. the number of different $T$'s, is
    upper bounded by ${n-1 \choose x}$.
  \item The number of different $v$ and $u_j$ is at most $n$ for each
    $j \in [1,x]$.
  \item Similarly to the argument for $T$, the number of different
    $S_j$ with $|\delta_H(S_j)| \leq k$ is at most
    $\sum_{t=1}^k {n-1 \choose t}$.
  \end{itemize}
  For $x \in [1,k]$, the number of different tuples $\alpha$ in
  $\Gamma$ with $|\delta_H(T)|=x$ is at most:
  \begin{equation*}
    n\cdot {n-1 \choose x} n \cdot n^x \cdot \left[\sum_{t=1}^k {n-1
      \choose t}\right]^x 
    = n^{O(k\cdot x)}.
  \end{equation*}
  Since $x \le k$, this gives us an upper bound of $n^{O(k^2)}$ on
  $|\Gamma|$.
\end{proof}

Suppose $\pi$ is the optimal mapping from $G$ to $H$ which yields a
mapping with cut distortion $k$ and distance distortion $\ell$ and
also certifies $z_{\alpha} = 1$ for some $\alpha$. Our recursive
algorithm does not necessarily obtain the same certificate as $\pi$
for $z_{\alpha} = 1$. So, before we show how to compute $z_{\alpha}$,
we examine certain properties of $z_{\alpha}$. In particular, we start
by proving that if both $\pi$ and $\gamma$ certify $z_{\alpha} = 1$ so
that $z_{\alpha,\pi}=z_{\alpha,\gamma} = 1$, then they not only match
on the boundary vertices but also on the cuts induced by boundary
edges.

\begin{lemma}\label{lem:z_alpha_properties}
  For $\alpha = (u,T,v,u_1,\dots,u_x,S_1,\dots,S_x)$, let $\pi$ and
  $\gamma$ be two mappings such that
  $z_{\alpha,\pi}=z_{\alpha,\gamma} = 1$. Then:
  
  \begin{enumerate}

  \item \label{item:z_alpha_1} $\pi(u) = \gamma(u)$.
  
  \item \label{item:z_alpha_2} $\pi(T_u^G) = \gamma(T_u^G)$.
  
  \item \label{item:z_alpha_3}
    For every boundary vertex $w$ of $T$(in $T$ with an incident
    edge in $\delta_H(T)$), $\pi^{-1}(w) =
    \gamma^{-1}(w)$. Equivalently, $\pi(u_j) = \gamma(u_j)$ for
    $j \in [1,x]$.
  \item \label{item:z_alpha_4} For every edge $e \in \delta_H(T)$,
    $\pi^{-1}(T_e^H) = \gamma^{-1}(T_e^H)$.
    
  \item \label{item:z_alpha_5} For every connected component $C$ in
    $H \setminus \delta_H(T)$, $\pi^{-1}(C) = \gamma^{-1}(C)$.
    
  \end{enumerate}  
    
\end{lemma}
\begin{proof}
  Items \ref{item:z_alpha_1}-\ref{item:z_alpha_4}.
  follow directly from the definition of $z_{\alpha,\pi}$. Consider a
  connected component $C$ in $H \setminus \delta_H(T)$. Let
  $\delta_H(C) = \{e_{i_1},\dots,e_{i_t}\}$. Without loss of
  generality, assume that $e_{i_1}$ is the edge closest to the root
  $r_H$. Then:
  \begin{equation*}
    \gamma^{-1}(C) = \gamma^{-1}(T_{e_{i_1}}) \setminus \cup_{j=2}^t \gamma^{-1}(T_{e_{i_j}}).
  \end{equation*}
  Item \ref{item:z_alpha_3} implies that
  $\gamma^{-1}(T_{e_{i_j}}) = \pi^{-1}(T_{e_{i_j}})$ for $j \in [1,t]$
  thus proving $\pi^{-1}(C) = \gamma^{-1}(C)$.
\end{proof}

The next lemma is somewhat like a converse of the previous lemma. It shows
that if we have a mapping $\pi$ such that $z_{\alpha,\pi} = 1$ and
another mapping $\gamma$ such that $\gamma$ matches with $\pi$ on the
subtree and the boundary vertices and edges, then
$z_{\alpha,\gamma} = 1$ as well.
\begin{lemma}\label{lem:changing_permutation} Let
  $\alpha = (u,T,v,u_1,\dots,u_x,S_1,\dots,S_x)$ and
  $\pi:V(G) \rightarrow V(H)$ be such that $z_{\alpha,\pi} = 1$. Let
  $\gamma:V(G) \rightarrow V(H)$ be such that
  \begin{enumerate}
  \item $\gamma(w) = \pi(w)$ for $w \in T_u^G$
  \item $\gamma(u_j) = \pi(u_j)$ for $j \in [1,x]$
  \item $\gamma(S_j) = \pi(S_j)$ for $j \in [1,x]$ ($\pi$ and $\gamma$
    may not be identical on every element of $S_j$)
  \end{enumerate}
  Then, $z_{\alpha,\gamma} = 1$.
\end{lemma}
\begin{proof}
  Follows immediately from
  definition~\ref{def:z_alpha}.
\end{proof}
Next we show how to change the optimal mapping such that it agrees
with the mapping found by our algorithm on the subtree and is still
optimal. Following lemma formalizes this statement:
\begin{lemma}\label{lem:combining_two_permutations}
  Let $\pi$ be a mapping such that
  $z_{\alpha,\pi}=z_{\alpha_1,\pi} = 1$ where
  \begin{center}
  $\alpha = (a,T,b,u_1,\dots,u_x,S_1,\dots,S_x) \in \Gamma$ and   $\alpha_1 =
  (a^1,T^1,b^1,u^1_1,\dots,u_{x_1}^1,S_1^1,\dots,S_{x_1}^1) \in
  \Gamma$
  \end{center}
 such that $a_1$ is a child of $a$ in $G$. Suppose,
  $\gamma_1$ is another mapping such that $z_{\alpha_1,\gamma_1} = 1$.
  Let $\zeta$ be defined as follows: $\zeta(u) = \gamma_1(u)$ for
  $u \in T_{a^1}^G$ and $\zeta(u) = \pi(u)$ otherwise. Then,
  $z_{\alpha,\zeta} = 1$.
\end{lemma}
\begin{proof}
  Items \ref{item:def-z-alpha-1}-\ref{item:def-z-alpha-4}
  of lemma \ref{def:z_alpha} can be easily verified
  for $z_{\alpha,\zeta}$. Here, we prove that the following property
  holds:
    $\forall e \in E[H[T]], |\delta_G(\zeta^{-1}(T_e^H))| \leq k.$
  %
  
  There are three possible cases, that we consider separately. \\
    \noindent \textbf{1.} $e = (b,b_1)$, $e \in \delta_H(T^1)$.  \\
      By definition,    $\zeta^{-1}(T_e^H) = \gamma_1^{-1}(T_e^H)$. Since,
    $z_{\alpha_1,\gamma_1} = 1$, we get
    $|\zeta^{-1}(T_e^H)| = |\gamma_1^{-1}(T_e^H)| \leq k$.
    
    \smallskip
    \noindent \textbf{2.} $e \in E[H[T\setminus T^1]]$. \\ By definition,
    $\zeta^{-1}(T_e^H) = \pi^{-1}(T_e^H)$. Since, $z_{\alpha,\pi} = 1$,,
    we get $|\zeta^{-1}(T_e^H)| = |\pi^{-1}(T_e^H)| \leq k$.
    
    \smallskip
    \noindent \textbf{3.} $e \in E[H[T^1]]$. \\ By definition,
    $\zeta^{-1}(T_e^H) = \gamma_1^{-1}(T_e^H)$ and since,
    $z_{\alpha_1,\gamma_1} = 1$, we get $|\zeta^{-1}(T_e^H)| \leq k$.
\end{proof}
The above lemmas show that even if we find mappings for subtrees which
are different from the optimal mappings, they can still be merged with
the optimal mappings. Hence, we may just find any of the mappings for
each $\alpha$ and then recursively combine mappings. The following
lemma states the result and shows how to make it constructive:
\begin{lemma}\label{lem:combining_permutation}
  Let $a \in V(G)$ be a vertex in $G$ with children $a^1, \dots,
  a^t$. Let 
  \begin{center}
  	$\alpha = (a,T,b,u_1,\dots,u_x,S_1,\dots,S_x)$ and 
  	  $\alpha_1 = (a^1,T^1,b^1,u_1^1,\dots,u_{x_1}^1,
  	S_1^1,\dots,S_{x_1}^1),\dots,\alpha_t =
  	(a^t,T^t,b^t,u_1^t,\dots,u_{x_t}^t, S_1^t,\dots,S_{x_t}^t)\in
  	\Gamma$.
  \end{center}
  Let $\pi:V(G) \rightarrow V(H)$ be a mapping such that
  $z_{\alpha,\pi}=z_{\alpha_1,\pi}=\dots=z_{\alpha_j,\pi} = 1$ for all
  $j \in [1,t]$.  If for each $j \in [1,t]$ there exists a mapping
   $\gamma_j : V(G) \rightarrow V(H)$ with $z_{\alpha_j,\gamma_j} =1$,
  then there exists $\pi':V(G) \rightarrow V(H)$ such that
  $z_{\alpha,\pi'} = 1$ and $\pi'(w) = \gamma_j(w)$ for
  $w \in T_{a_j}^G$ where $j \in [1,t]$. Moreover, given
  $\{\gamma_j \mid j \in [1,t]\}$, such $\pi'$ can be found in time
  $poly(n)$.
\end{lemma}
\begin{proof}
  Let $\zeta:V(G) \rightarrow V(H)$ such that $\zeta(w) = \gamma_j(w)$
  for $w \in T_{a_j}^G$ where $j \in [1,t]$ and $\zeta(w) = \pi(w)$
  otherwise. By lemma~\ref{lem:combining_two_permutations},
  $z_{\alpha,\zeta} = 1$.

  \noindent {\bf Construction of $\pi'$:} Let $\pi'(w) = \gamma_j(w) $
  for $w \in T_{a_j}^G, j \in [1,t]$ and $\pi'(a) =b$. For
  $w \not\in T_a^G$, define $\pi'$ such that $\pi'(S_j) = T_j$ for
  $j \in [1,x]$.   Setting $\pi = \zeta, \gamma = \pi'$ in lemma~\ref{lem:changing_permutation}, we get $z_{\alpha,\pi'} = 1$.  Easy
  to see that $\pi'$ is constructed in polynomial time.
\end{proof}
Lemma \ref{lem:combining_permutation} suggests that we can recursively
compute $z_{\alpha}$. Namely, we can show the following:
\begin{lemma}\label{lem:computing_z_alpha}
  There exists an algorithm with running time $poly(n,|\Gamma|^d)$
  which calculates $z_{\alpha}$ for each $\alpha \in
  \Gamma$. Additionally if $z_{\alpha} = 1$, it also computes
  $\pi_{\alpha}$ such that $z_{\alpha,\pi_{\alpha}} = 1$.
\end{lemma}
\begin{proof}
  Consider $\alpha = (a,T,b,u_1,\dots,u_x,S_1,\dots,S_x)\in \Gamma$
  with $z_{\alpha} = 1$ and $\pi:V(G) \rightarrow V(H)$ be the mapping
  such that $z_{\alpha,\pi}=1$. Let the children of $a$ be
  $a^1,\dots,a^t$.
  \begin{claim}
    For
    $j \in [1,t],\exists \alpha_j =
    (a^j,T^j,b^j,u_j^1,\dots,u_{x_j}^1, S_1^1,\dots,S_{x_j}^1)\in
    \Gamma$ such that $z_{\alpha_j,\pi} = 1$.
  \end{claim}
  To construct a mapping $\pi'$ such that $z_{\alpha,\pi'} = 1$, we
  guess $\alpha_1,\dots,\alpha_t$ and use lemma
  \ref{lem:combining_permutation} to construct such a mapping. It
  requires mapping $\gamma_j$ such that $z_{\alpha_j,\gamma_j} = 1$
  which can be assumed to be constructed recursively. The number of
  choices of $\alpha_1,\dots,\alpha_t$ is upper bounded by
  $|\Gamma|^t$ which is upper bounded by $|\Gamma|^d$, as the degree
  of any vertex is at most $d$. For any such choice, algorithm in lemma~\ref{lem:combining_permutation} runs in time $poly(n)$. Hence,
  computing $z_{\alpha}$ takes $|\Gamma|^d poly(n)$ time for each
  $\alpha$ and $|\Gamma|^{d+1} poly(n) = poly(n,|\Gamma|^d)$ time for
  all $\alpha \in \Gamma$.

  If $z_{\alpha} = 0$, then for any of the mappings $\pi'$ considered
  above has $z_{\alpha,\pi'} = 0$. This can be checked in $poly(n)$
  time for each $\pi'$ (proposition~\ref{prop:z_alpha_pi}).
\end{proof}

\begin{proof}(of theorem~\ref{thm:cuts_distances_algo}) Let $\pi$ be a mapping
  $\pi: V(G) \rightarrow V(H)$ such that:

  (a) For all $(u,v) \in E(G)$, $d_H(\pi(u),\pi(v)) \leq \ell$
  
  (b) For all $(u,v) \in E(H)$,
    $d_G(\pi^{-1}(u),\pi^{-1}(v)) \leq \ell$
  
  (c) For $S \subset V(G)$ s.t. $|\delta_G(S)| = 1$,
    $|\delta_H(\pi(S))| \leq k$.
  
  (d) For $S \subset V(H)$ s.t. $|\delta_H(S)| = 1$,
    $|\delta_G(\pi^{-1}(S))| \leq k$.

  First, we start by guessing the roots of $G$ and $H$ and define
  $\Gamma$. Then, using  lemma \ref{lem:computing_z_alpha}, we can calculate
  $z_{\alpha}$ for $\alpha \in \Gamma$. It does not give us a mapping
  $\pi'$ satisfying the conditions above since for
  $\alpha = (u,T,v,u_1,\dots,u_x,S_1,\dots,S_x) \in \Gamma$, we have
  $u \neq r_G$. However, a proof almost identical to that of
  lemma~\ref{lem:computing_z_alpha} works here as well. Assume $r_G$ has
  children $a_1,\dots,a_t$. Then there exists
  $\alpha_j = (a^j,T^j,b^j,u_j^1,\dots,u_{x_j}^1,
  S_1^1,\dots,S_{x_j}^1)\in \Gamma$ such that $z_{\alpha_j,\pi} =
  1$. Then, similarly to the proof of lemma~\ref{lem:computing_z_alpha} we
  can guess $\alpha_j,j \in [1,t]$ and compute $\pi'$ in time
  $poly(n)\cdot |\Gamma|^d$, which satisfies the conditions described
  above.
\end{proof}

\section{Final Remarks} \label{sec:future}
%
%

%
From an algebraic standpoint, the problems we considered in this work have natural generalizations to pairs of positive definite matrices $(A,B)$, and the corresponding eigenvalue problem $Ax = \lambda P^T B Px$. SGD generalizes to minimizing the maximum eigenvalue, and SGRI generalizes to finding the permutation $P$ that minimizes the condition number $\kappa(A,B)$. 
But the problem appears to be much harder in some sense: one can construct `pathological' examples of $A$ and $B$ with just two distinct eigenspaces that are nearly identical, but different enough to cause unbounded condition numbers due to the eigenvalue gap. This makes implausible the existence of non-trivial subexponential time algorithms for the general case. 

On the other hand, besides their potential for applications, Laplacians seem to offer an interesting mathematical ground with a wealth of open problems. In this paper we presented the first algorithmic result, for unweighted trees. The algorithm is admittedly complicated, but it can at least be viewed as an indication of algorithmic potential, as we are not aware of any fact that would preclude a $\kappa^2$-approximation for general graphs. To make such algorithmic progress, we would likely have to give up on the combinatorial interpretations of the condition number, and use deeper spectral properties of Laplacians.

\appendix

\newpage
\section{Bounding Condition Numbers: Sufficient and Necessary Properties}~\label{sec:necessary_sufficient}
We first describe the combinatorial implications of a bounded condition number. 

\begin{lemma} [cuts] \label{lem:cut-nec}
  Let $G$ and be two graphs such that for all $x$,   
  $\beta \qf(H,x) \leq \qf(G,x) \leq \gamma \qf(H,x)$.
  Then, for any $S\subseteq V$, we have
  $\beta \delta_H(S) \leq \delta_G(S) \leq \gamma \delta_H (S)$.
\end{lemma}
\begin{proof}
  Let $x_S$ be the indicator vectors with $x_u=1$ for $u\in S$
  and $x_u=0$ for $u \not \in S$. We have $\qf(G,x_S) = \delta_G(S)$ and similarly for $H$. 
  The proof follows by  substituting  $x_S$ into $x$ in the given inequalities. 
\end{proof}
\begin{lemma} [distances] \label{lem:dist-nec}
  Let $G$ and $H$ be two trees such that for all $x$,   
  $\beta \qf(H,x) \leq \qf(G,x) \leq \gamma \qf(H,x)$
  Then
  $\beta \le \frac{d_H(u,v)}{d_G(u,v)} \le \gamma$ for any pair of
  nodes $u\neq v \in V$.
\end{lemma}
\begin{remark}
  If we replace the shortest path metric in lemma~\ref{lem:dist-nec} with
  that of the resistance distance (the resistance between two
  equivalent points on an electrical network corresponding to $G$);
  then it holds for any graph~\cite{doylesnell00}.
\end{remark}
\begin{proof}
 Assuming that for all $x$ we have $x^T L_G x \leq \gamma x^T L_H x$, 
 it can be shown that we also have  $x^T L_H^{\dagger} x \leq \gamma x^T L_G^{\dagger} x$~\cite{support03boman}, where $L_G^{\dagger}$ is the pseudo-inverse of $L$.  
 It then suffices to prove that $x^T L_G^{\dagger} x = d_G(u, v)$ when $G$ is a
 tree.  Given then a pair $u, v$,   consider the vector $x=[x_a]_{a\in V}$ where $$x_a =
  \begin{cases}
    +1 & \mbox{if $u = a$,} \\
    -1 & \mbox{if $v = a$,} \\
    0 & \mbox{otherwise.}
  \end{cases}$$
  Consider the path $\pi$ from $u$ to $v$ in $G$. For any node $a$, we
  will abuse the notation and use $\pi(a)$ to denote the closest
  ancestor of $a$ along the path $\pi$, i.e., the first node in $\pi$
  one encounters along the walk from $a$ to (say) $u$. Now consider
  the following vector $y=[y_a]_{a\in V}$ where $y_a$ is equal to the
  distance of $\pi(a)$ to $u$.  By construction and the graph being a
  tree, $y^T L y$ counts the number of edges along the path $\pi$.
  Since the vector $x$ we chose is orthogonal to all $1$'s vector,
  whose span is equal to the kernel of $L$, it suffices to prove that
  $x = L y$.
  For any $a\in V$, we have:
  \[
    (L y)_a 
    = \sum_{b \in N(a)} (y_a - y_b)
    =  \sum_{b\in N(a) \cap \pi} (y_a - y_b).
  \]
  There are four cases, three of which are trivial:
  \begin{itemize}
  \item If $a=u$, then this sum is $-1$. 
  \item If $a=v$, then this sum is $+1$. 
  \item If $a \notin \pi$, the sum becomes $0$.
  \item Finally if $a \in \pi\setminus\{u,v\}$, with its successor and
    predecessor along $\pi$ being $s$ and $t$, respectively; then:
    \[
      (L y)_a 
      = (y_a - y_s) + (y_a - y_t)
      = 1 - 1 = 0. 
    \]    
  \end{itemize}
  Therefore $L y = x = e_u - e_v$ as expected.
\end{proof}
%
%
It turns out that cuts and distances are also sufficient to
get an upper bound on the support numbers and thus on the condition number~\cite{GuatteryM00}.
\begin{lemma}
  \label{thm:dila-cong-suff}
  Given two graphs $G$ and $H$  if there exist a flow $f$
  in $H$ such that the following  conditions are true:
  \begin{itemize}
  \item For each edge $(u,v) \in E_G$, $f$ routes one unit of flow
    from $u$ to $v$ in $H$ over paths of length at most $\alpha$.
  \item Flow $f$ has congestion at most $\beta$ in $H$.
  \end{itemize}
  Then for all $x$, $\qf(G,x) \le \alpha\beta \qf(H,x)$, i.e. $\sigma(G,H) \leq \alpha\beta$.
\end{lemma}
\begin{proof}
  First we will prove that if such $f_2$ exists, then
  $L \preceq \alpha\beta M$. For any vector $x\in \mathbb{R}^n$:
  \begin{eqnarray*}
     \qf(G,x)
     & = &
        \sum_{uv\in E(G)}
        (x_u - x_v)^2 
        = \sum_{uv\in E(G)}
        \left[\sum_{ab \in f(uv)} (x_a - x_b)\right]^2 \\
     & \le &
          \sum_{uv\in E(G)}
          |f(uv)| 
          \sum_{ab \in f(uv)} (x_a - x_b)^2 \\
    & \le & \alpha \sum_{ab\in E(H)}
          \left|
          \left\{
          uv \in E(G)
          \big| 
          ab \in f(uv)
          \right\}
          \right|
          (x_a - x_b)^2 \\
        &  \le & \alpha \beta \sum_{ab\in E(H)}
          (x_a - x_b)^2 \\
    & =& \alpha \beta \qf(H,x).
  \end{eqnarray*}
  Here $f(uv)$ denotes the edges along the path
  assigned to the demand pair $u$ and $v$.  	
\end{proof}

\begin{theorem}
  \label{thm:cuts-edges-trees}
  Given trees $G$ and $H$ with Laplacian matrices $L$ and
  $M$, respectively on the same set of nodes, let $k$ and $\ell$ be the minimum values for
  which:
  \begin{enumerate}
  \item (Stretch) For any edge $(u,v)$ of $G$, 
   $d_H(u,v) \le \ell$ 
  \item  (Cut) For any edge $(u,v)$ of $H$, 
    $\Big|\delta_G\big[T_u^G(v)\big]\Big| \le k$
  \end{enumerate} 
  Then $\max\{k,\ell\} \le \sigma(G, H) \le k \ell$.
  Here, for two nodes $u$ and $v$ of a tree $G$, $T_u^G(v)$ denotes
  the subtree of $G$ at $v$ when $u$ is identified as the root and
  $\delta_H\big[T_u^G(v)\big]$ denotes the corresponding set of edges
  in $H$ that cross the cut $(A,A')$, where $A$ contains the set of
  nodes of $T_u^G(v)$.
\end{theorem}
\begin{proof}
  The lower bound, $\max\{k,\ell\} \le \sigma (G, H)$, follows easily
  from lemmas~\ref{lem:cut-nec} and~\ref{lem:dist-nec}.
  In order to prove the upper bound, we will consider the natural
  multicommodity flow $f$ with demand graph $G$ and capacity
  graph $H$.  For each edge
  $(u,v)$ of $G$, $f$ has a unit flow along the unique path between
  $u$ and $v$ in $H$.   By the stretch condition, $f$ routes flows through
  paths of length at most $\ell$. Now we will bound the congestion.
  Consider any edge $e=(u,v)$ of $H$. Let $A$ be the connected
  component of $H$ containing $u$ after removing $e$.  Observe that
  this is the same as subtree of $H$ at $u$ when $v$ is identified as
  the root, $A = T_v^H(u)$. If $(s,t)$ is an edge of $G$ which sends
  flow across $e$, then $s$ and $t$ should lie in different connected
  components of $G$ after the removal of $e$. If we assume, without
  loss of generality, that $s \in A$; then $t \in \overline{A}$. So
  the congestion of $e$ is equal to the number of edges of $G$
  crossing $A$, $|\delta_H(A)| = |\delta_G[T_u^H(v)]| \le k$. 
   Thus we can invoke \ref{thm:dila-cong-suff} and
  obtain the desired upper bound, $\sigma (G, H) \le k \ell$.
\end{proof}



\end{document}